\documentclass[aps,twocolumn,superscriptaddress,prl]{revtex4-1}

\usepackage{bm}
\usepackage{graphicx}
\usepackage{subfigure}
\usepackage{epstopdf}
\usepackage{siunitx}

\usepackage{color}

\def\bem#1{\begin{mathletters}\label{#1}}
\def\eml{\end{mathletters}}

\def\ket#1{{|#1\rangle}}

\def\4#1{{\boldsymbol{#1}}}
\def\8#1{{\widetilde{#1}}}

\begin{document}

\title{Purcell-enhanced optical spin readout of Nitrogen-Vacancy centers in diamond}

\author{S. A. Wolf}
\affiliation{The Racah Institute of Physics, The Hebrew University of Jerusalem, Jerusalem 91904, Israel}
\affiliation{The Center for Nanoscience and Nanotechnology, The Hebrew University of Jerusalem, Jerusalem 91904, Israel}

\author{I. Rosenberg}
\affiliation{The Racah Institute of Physics, The Hebrew University of Jerusalem, Jerusalem 91904, Israel}
\affiliation{The Center for Nanoscience and Nanotechnology, The Hebrew University of Jerusalem, Jerusalem 91904, Israel}

\author{R. Rapaport}
\affiliation{The Racah Institute of Physics, The Hebrew University of Jerusalem, Jerusalem 91904, Israel}
\affiliation{The Center for Nanoscience and Nanotechnology, The Hebrew University of Jerusalem, Jerusalem 91904, Israel}
\affiliation{Dept. of Applied Physics, Rachel and Selim School of Engineering, Hebrew University, Jerusalem 91904, Israel}

\author{N. Bar-Gill}
\email{bargill@phys.huji.ac.il}
\thanks{Corresponding author.}
\affiliation{Dept. of Applied Physics, Rachel and Selim School of Engineering, Hebrew University, Jerusalem 91904, Israel}
\affiliation{The Racah Institute of Physics, The Hebrew University of Jerusalem, Jerusalem 91904, Israel}
\affiliation{The Center for Nanoscience and Nanotechnology, The Hebrew University of Jerusalem, Jerusalem 91904, Israel}

\begin{abstract}
Nitrogen-Vacancy (NV) color centers in diamond have emerged as promising quantum solid-state systems, with applications ranging from quantum information processing to magnetic sensing. One of the most useful properties of NVs is the ability to read their ground-state spin projection optically at room temperature. This work provides a theoretical analysis of Purcell enhanced NV optical coupling, through which we find optimal parameters for maximal Signal to Noise Ratio (SNR) of the optical spin-state readout. We conclude that a combined increase in spontaneous emission (through Purcell enhancement) and in optical excitation could significantly increase the readout SNR. 
\end{abstract}

\maketitle

Isolated solid-state quantum systems, such as quantum dots, NV centers in diamond and other point-like defects, coupled to photonic degrees-of-freedom, could serve as single-photon sources and quantum nodes \cite{Holmes2014,Hucul2014}. This hybrid approach to quantum networks and quantum information processing architectures combines the benefits of solid-state storage and manipulation with photonic communications and distribution.

A commonly encountered problem is the limited coupling efficiency between the quantum system and light, which hinders the scalability of the proposed quantum networks and the rate of information transfer. Several approaches have been suggested to enahnce the coupling efficiency including optical cavities and waveguides \cite{Yoshie2004,Gao2012,Burek2014,Li2015}, as well as nanofabricated plasmonic and dielectric structures \cite{Bulu11,Maletinsky2014}. Generally speaking, optical antennas selectively enhance the optical coupling to a specific resonant mode, resulting in modified spontaneous emission, called Purcell Factor (PF), and in high directionality of this emission, leading to a better photon collection efficiency \cite{livneh11,haratz14}. The resulting enhancement could have a major impact on the applicability of these systems, advancing toward the goal of efficient solid-state/light interfaces and potentially achieving single-shot readout of the quantum state of the system.

However, in certain cases the modified behavior of the quantum system in the presence of the resonant antenna (and given the resulting Purcell enhancement of its spontaneous emission) might degrade its usefulness as a quantum emitter/node. Specifically, for the case of NV centers in diamond, the optical readout of the defect's spin state relies on the spin-dependent branching ratio between radiative and non-radiative decay paths. By enhancing spontaneous emission, optical antennas may diminish the difference in fluorescence emitted by different spin states and hence reduce readout contrast.
 
\begin{figure}[tbh]
\begin{center}
\subfigure[]{
\includegraphics[width=0.85 \linewidth]{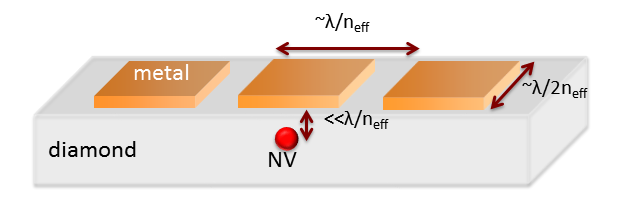}}
\subfigure[]{
\includegraphics[trim = 3mm 1mm 0mm 3mm, clip, width=0.46 \linewidth]{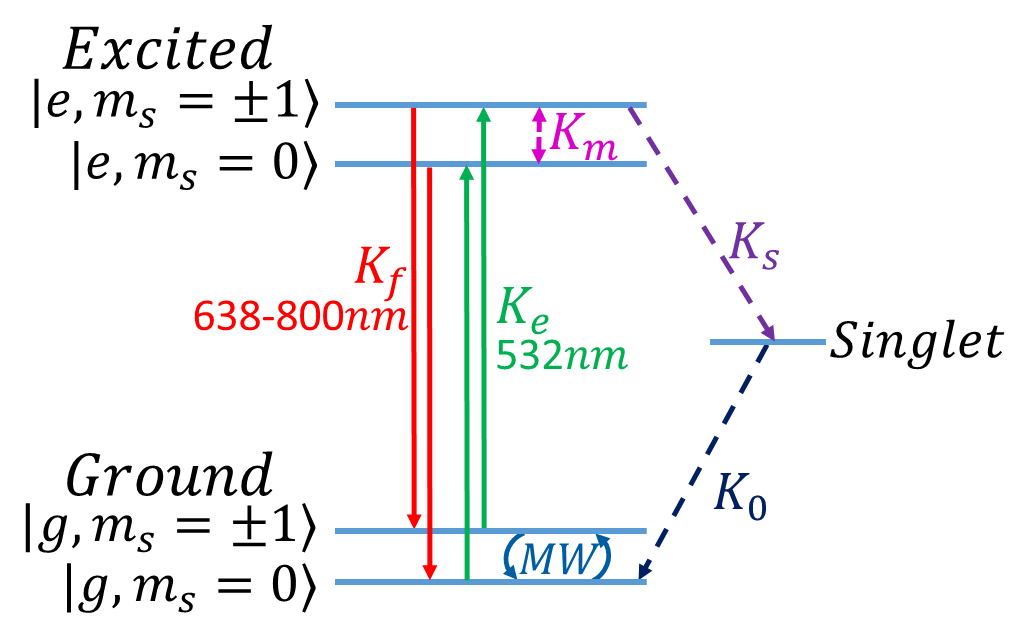}}
\subfigure[]{
\includegraphics[trim = 3mm 1mm 0mm 3mm, clip, width=0.46 \linewidth]{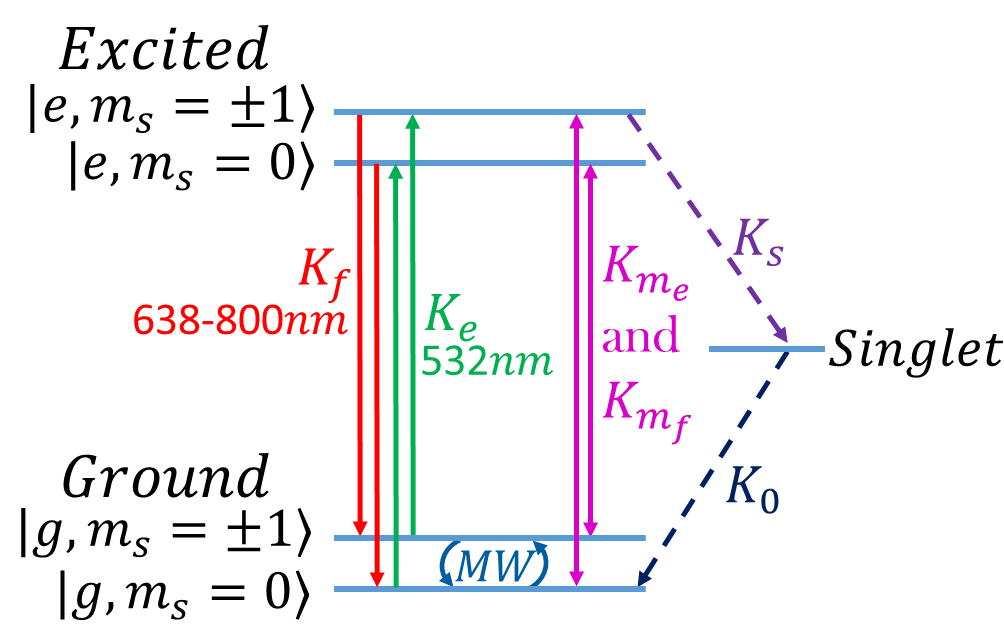}}
\protect\caption{NV coupled to an optical antenna, energy levels and allowed transition. (a) A general schematic example of a plasmonic antenna designed to yield a Purcell enhancement of the radiative rate and to direct the emission. (b) Energy levels and allowed transitions for NV center with non-radiative spin-mixing transitions in the excited state. $K_{e}$ is the spin-preserving excitation rate (with wavelength $=532nm$), $K_f$ is the spin-preserving radiative decay rate, $K_s$ is the non-radiative decay rate to the singlet and $K_0$ is the non-radiative decay rate from the singlet. $K_m$ is the non-radiative non-spin-preserving rate from $|e,m_{s}=0\rangle$ to $|e,m_{s}=\pm 1\rangle$ and vice versa. $P_{g,1}$ and $P_{e,1}$ include spin projections $m_s=\pm 1$.
(c) Energy levels and allowed transitions in NV center with radiatively-activated spin-mixing transitions. $K_{m_e}$ and $K_{m_f}$ are the spin-mixing radiative excitation and decay rates. Spin-mixing of radiative origin [replacing $K_m$ in (b)] is described through $K_{m_e}$ and $K_{m_f}$ rates that scale with the Purcell Factor (PF).} \label{fig:rates}
\end{center}
\end{figure}

In this work we consider an NV center coupled to an antenna [Fig. \ref{fig:rates}(a)] and investigate its effect on spin-state readout. A leading candidate for the realization of such an antenna is a plasmonic structure, since it can accommodate a wide resonance to cover a large part of the broad NV phonon-sideband emission, yet still allow for large Purcell factors (PFs) due to its very small optical mode volume \cite{belacel13,Bulu11}. We focus on the interplay between the readout Signal-to-Noise Ratio (SNR) and the Purcell enhancement of radiative decay for this system, and find experimentally feasible optimal parameters that significantly enhance both optical coupling and readout fidelity. We note that this result relies on the assumption that spin-mixing terms are nonradiative in nature \cite{jelezko2004,Goldman14}, since otherwise the improvement in SNR will be limited (as further discussed below).

The NV center consists of a substitutional nitrogen atom and a vacancy occupying adjacent lattice sites in the diamond crystal. The electronic ground state is a spin triplet, in which the $m_\mathrm{s}=0$ and $m_\mathrm{s}=\pm 1$ sublevels experience a $\sim 2.87$ GHz zero-field splitting [Fig.\,\ref{fig:rates}(b)], while a static magnetic field can further split the $\pm 1$ sublevels to create an effective two-level system. The NV spin can be initialized with optical excitation, detected via state-dependent fluorescence intensity, and coherently manipulated using microwaves \cite{childress2006}.

The optical readout of the NV center relies on excitation and fluorescence intensity measurements. Experimentally, the spin state of the NV is extracted from the emitted fluorescence intensity under optical (green) excitation at a wavelength of $532$ nm. The excitation laser is switched on for spin-state readout, and simultaneously fluorescence (in a red phonon-sideband between $650-800$ nm) is collected during a readout time window $T$. 
The dynamics of the NV under optical excitation are determined by rate equations, based on the transition rates between different energy levels as depicted in Fig. \ref{fig:rates}(b). 

The ability to accurately measure the state of an NV center relies on the difference in the number of photons emitted during measurements of the two spin projections in the ground state, $|g,m_{s}=0\rangle$ (ground state with spin projection $m_{s}=0$) and $|g,m_{s}=\pm 1\rangle$ (ground state with spin projection $m_{s}=\pm 1$), and is quantified by the measurement SNR defined below. Let $N_{0}$ and $N_{1}$ be the random variables representing the number of photons emitted during a measurement of $|g,m_{s}=0\rangle$ and $|g,m_{s}=\pm 1\rangle$ respectively. $N_{0}$ is the number of photons emitted during a measurement of the NV spin state $|g,m_{s}=0\rangle$ and $N_1$ is the number of photons emitted during a measurement of the NV spin state $|g,m_{s}=\pm 1\rangle$. Since the number of photons emitted from the laser is Poisson distributed, we will assume that $N_{0}$ and $N_{1}$ are both Poisson distributed as well. Therefore the difference $\Delta=N_{0}-N_{1}$ is Skellam distributed \cite{PoissonAlzaid10}. Denoting $n_{0}$ and $n_{1}$ to be the expected values of $N_{0}$ and $N_{1}$ respectively, the expected value and variance of the difference, $\Delta$, are 
\begin{eqnarray}
E \left( \Delta \right) &=& n_{0}-n_{1}, \nonumber \\
\sigma^2 \left( \Delta \right) &=& n_{0}+n_{1}. 
\end{eqnarray}
The SNR of the Skellam distribution, defined as $SNR=\frac{E\left(\Delta\right)}{\sigma\left(\Delta\right)}$, is thus \cite{Steiner10}
\begin{equation}
SNR=\frac{n_{0}-n_{1}}{\sqrt{n_{0}+n_{1}}}. \label{Eq:SNR}
\end{equation}

The importance of the SNR stems also from its relation to NV state estimation. An (unexcited) NV state can be written as $|\psi_{r}\rangle=r|g,m_{s}=0\rangle+\left(1-r\right)|g,m_{s}=\pm 1\rangle$, where $0\leq r\leq1$, is the fraction of ground state with spin projection $m_{s}=0$ (this definition could relate to a mixed-state or to a pure state, and, without loss of generality, $r$ is taken to be real for simplicity). In the Supplementary Material \cite{SuppMat} we derive the maximum likelihood estimation of $r$ and show that the minimization of the resulting estimation error is equivalent to maximizing the SNR in Eq. \ref{Eq:SNR}.

In order to better understand the role of each of the system's rates in the SNR, we present a theoretical analysis of the NV measurement process. We rely on the following assumptions: 
\begin{enumerate}
\item The orbital mixing rate of the excited state is much larger than the decay rate \cite{Goldman14} (this causes orbital averaging, essentially resulting in a single spin-triplet in the excited state).
\item The radiative decay rate of the excited states is independent of the spin projection \cite{manson2006}. 
\item The excitation rate from both ground states is the same \cite{manson2006}. 
\item The transition from the excited state with $m_{s}=0$ to the singlet is negligible (since it is 4 orders of magnitude smaller than from $m_{s}=\pm 1$)  \cite{Young2009,manson2006}. 
\item Both ground state spin sublevels $m_s=-1$ and $m_s=+1$ have the same dynamics. 
\item The singlet state decays to $|g,m_{s}=0\rangle$\cite{manson2006}. 
\item The spin-mixing rate is the same between $\pm 1 \leftrightarrow 0$.
\end{enumerate}

Given these assumptions, the rate equations that govern the transitions in the NV center are:

\begin{eqnarray}
\dot{P}_{g,0} & = & -K_{e}P_{g,0}+K_{f}P_{e,0}+K_{0}P_{s} \nonumber \\
\dot{P}_{g,1} & = & -K_{e}P_{g,1}+K_{f}P_{e,1} \nonumber \\
\dot{P}_{e,0} & = & K_{e}P_{g,0}-(K_{f}+2K_{m})P_{e,0}+K_{m}P_{e,1} \nonumber \\
\dot{P}_{e,1} & = & K_{e}P_{g,1}-(K_{f}+K_{s}+K_{m})P_{e,1}+2K_{m}P_{e,0} \nonumber \\
\dot{P}_{s} & = & K_{s}P_{e,1}-K_{0}P_{s}, \label{Eq:rates}
\end{eqnarray}
where $P_{g,0}$ and $P_{g,1}$ are the populations of the ground state with $m_{s}=0$ and $m_{s}=\pm 1$ respectively, $P_{e,0}$ and $P_{e,1}$ are the populations of the excited state with $m_{s}=0$ and $m_{s}=\pm 1$ respectively, and $P_{s}$ is the population of the singlet (Fig. \ref{fig:rates}). $K_{e}$ is the excitation rate (spin-state preserving), $K_{f}$ is the radiative decay rate (spin-state preserving), $K_{0}$ is the decay rate from the singlet to $|g,m_{s}=0\rangle$, and $K_{s}$ is the decay rate from $|e,m_{s}=1\rangle$ (excited state with spin projection $m_{s}=\pm 1$) to the singlet. $K_{m}$ denotes small spin mixing transitions assumed to be related to phononic coupling in the excited state [Fig. \ref{fig:rates}(b)] \cite{Goldman14}. The sum of the populations is normalized to be one, $P_{g,0}+P_{g,1}+P_{e,0}+P_{e,1}+P_{s}=1$. In the case of spin-mixing rates of radiative origin [Fig. \ref{fig:rates}(c)] the rate equations will be slightly different [Eq. (1) in \cite{SuppMat}], with spin-mixing rates which scale with PF and $K_e$. In the main part of the paper we assume non-radiative spin mixing, i.e. constant mixing rate, $K_{m}$, between the excited states [Fig. \ref{fig:rates}(b)].

\begin{figure}[tbh]
\begin{center}
\begin{subfigure}[]{
\includegraphics[trim = 1mm 1mm 10mm 6mm, clip, width=0.47 \linewidth]{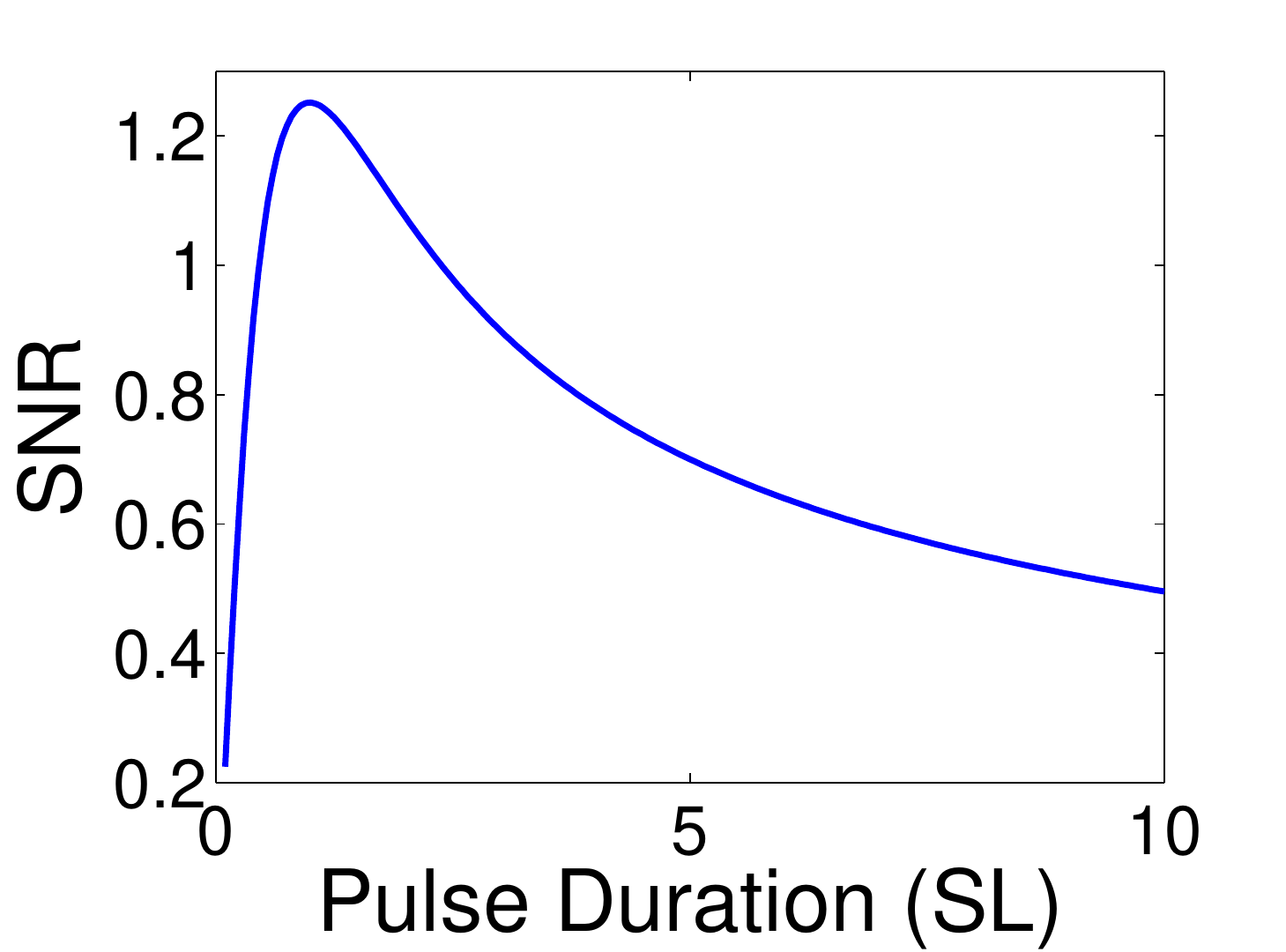}}
\end{subfigure}
\begin{subfigure}[]{
\includegraphics[trim = 1mm 1mm 5mm 5mm, clip, width=0.47 \linewidth]{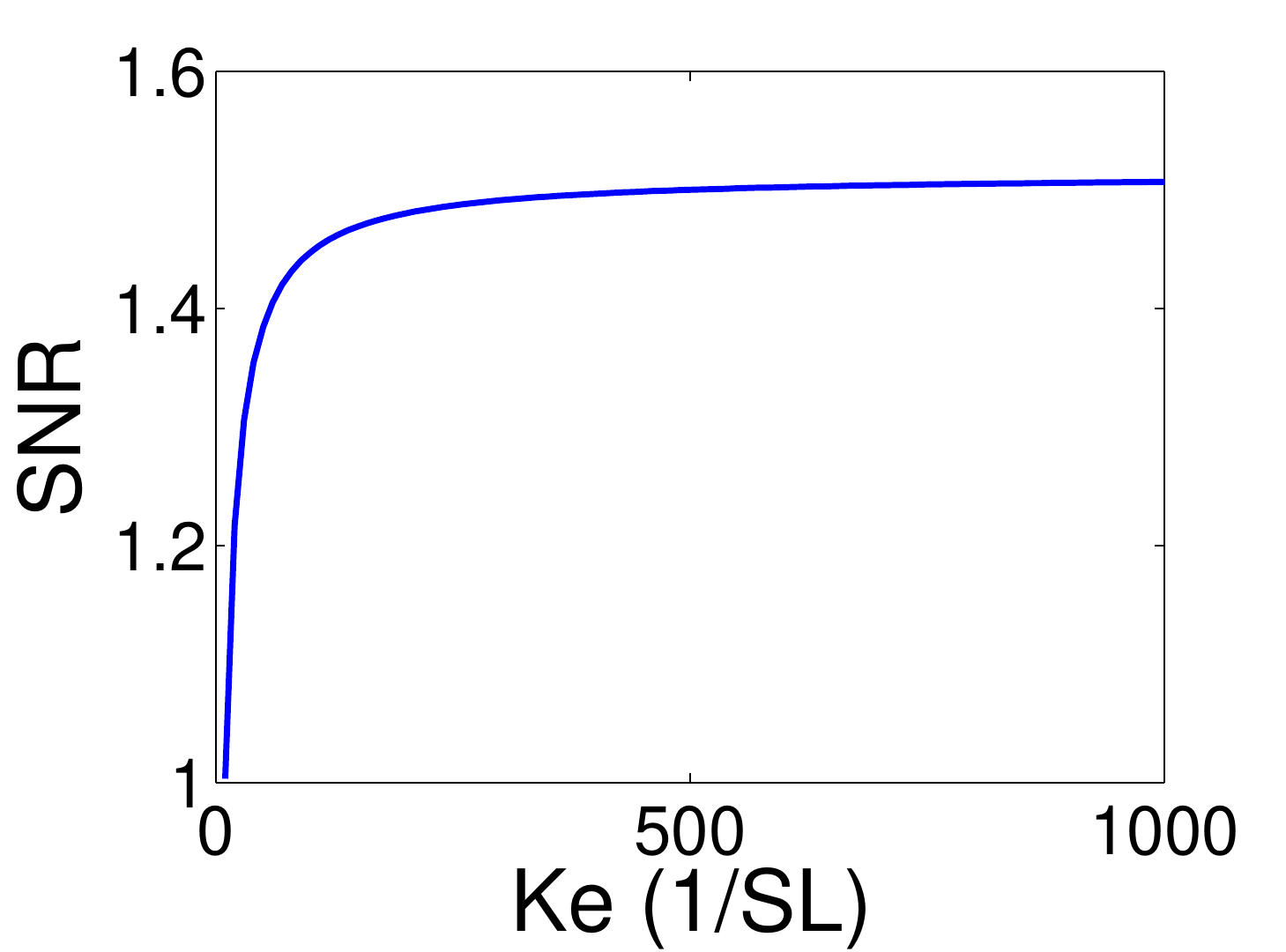}}
\end{subfigure}
\protect\caption{(a) SNR as a function of the pulse duration, T, for typical rates (see text). The optimal pulse duration is $T=0.9911 [SL]$, at which  the maximum value of $SNR=1.2516$ is reached. (b) SNR as a function of the excitation rate, $K_{e}$, for typical rates and pulse duration $T=1 [SL]$. } \label{fig:T_Ke}
\end{center}
\end{figure}

In the following analysis we used dimensionless variables by normalizing the rates and durations by the lifetime of the singlet state ($\approx 300$ ns), which will be referred to as the Singlet Lifetime (SL). Following this normalization the typical dimensionless NV rates are: $K_{f_0}=\frac{300}{13} \approx 23.077 [\frac{1}{SL}]$ (the unmodified radiative decay rate), $K_{s}=\left(\frac{300}{7.8}-\frac{300}{13}\right) \approx 15.3846 [\frac{1}{SL}]$, $K_{0}=1 [\frac{1}{SL}]$, $K_{m}=0.0404K_{f_0} \approx 0.9323 [\frac{1}{SL}]$ \cite{manson2006}. We also take $K_e = K_{f_0}$ as the typical excitation rate. 

For a given set of rates, the SNR, derived from the solution of the rate equations (Eq. \ref{Eq:rates}), peaks at an optimal pulse duration, as is evident from Figure \ref{fig:T_Ke}(a) for the typical rates above. Moreover, for a given pulse duration the SNR increases as a function of the excitation rate, $K_{e}$, and at large values of $K_{e}$ the SNR saturates asymptotically [Fig. \ref{fig:T_Ke}(b)]. Both phenomena are in agreement with previously published experimental results \cite{Doherty13,manson2006,Steiner10}, thereby supporting our analysis and assumptions. For the typical NV rates the optimal pulse duration is $T \approx 1 [SL]$ [Fig. \ref{fig:T_Ke}(a)], which results in the maximum $SNR=1.2516$.

\begin{figure}[tbh]
	\begin{center}
		\begin{subfigure}[ ]{
				\includegraphics[trim = 1mm 0mm 10mm 6mm, clip, width=0.47 \columnwidth]{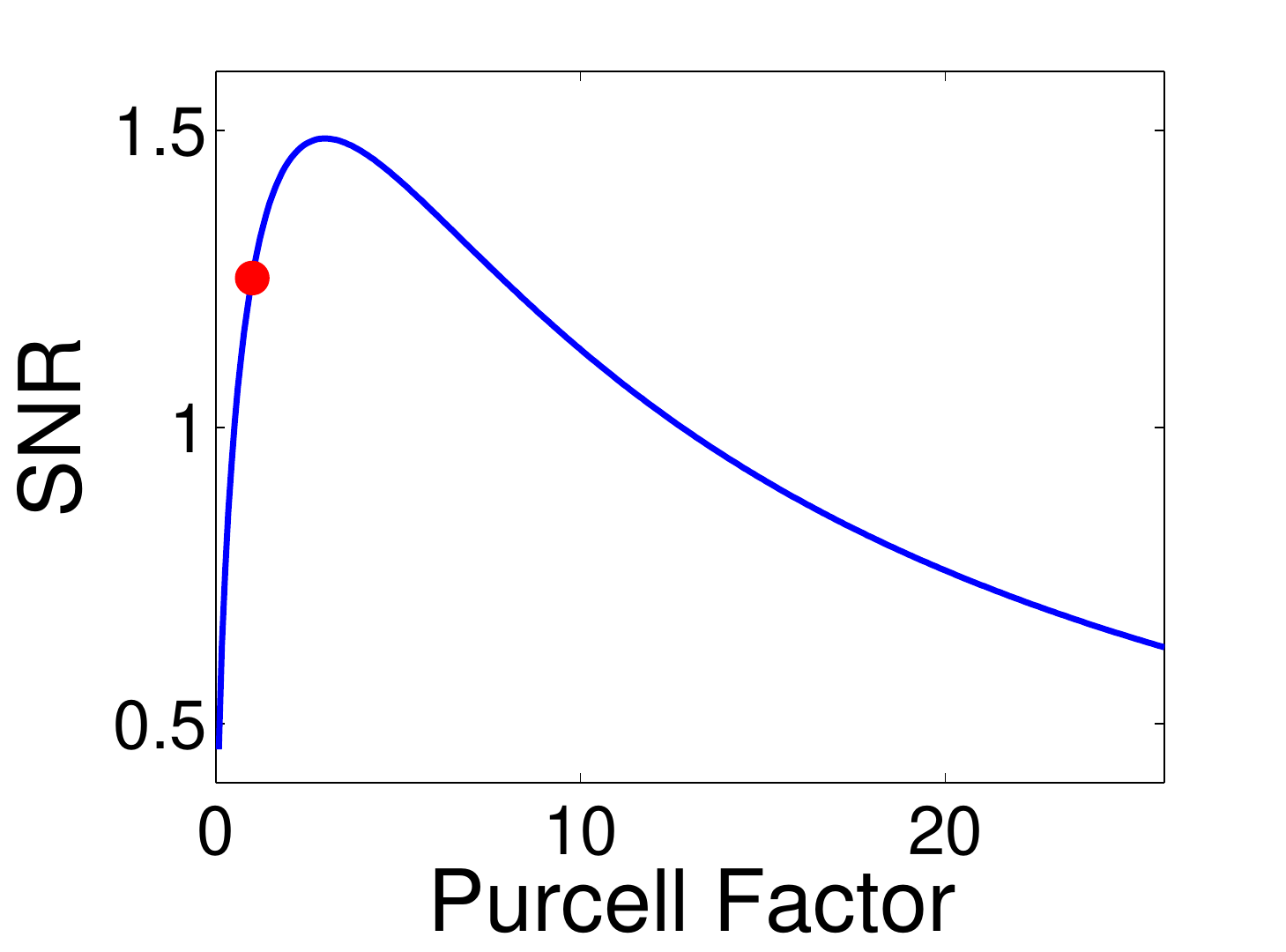}}
		\end{subfigure}
		\begin{subfigure}[ ]{
				\includegraphics[trim = 2mm 5mm 1mm 6mm, clip, width=0.47 \columnwidth]{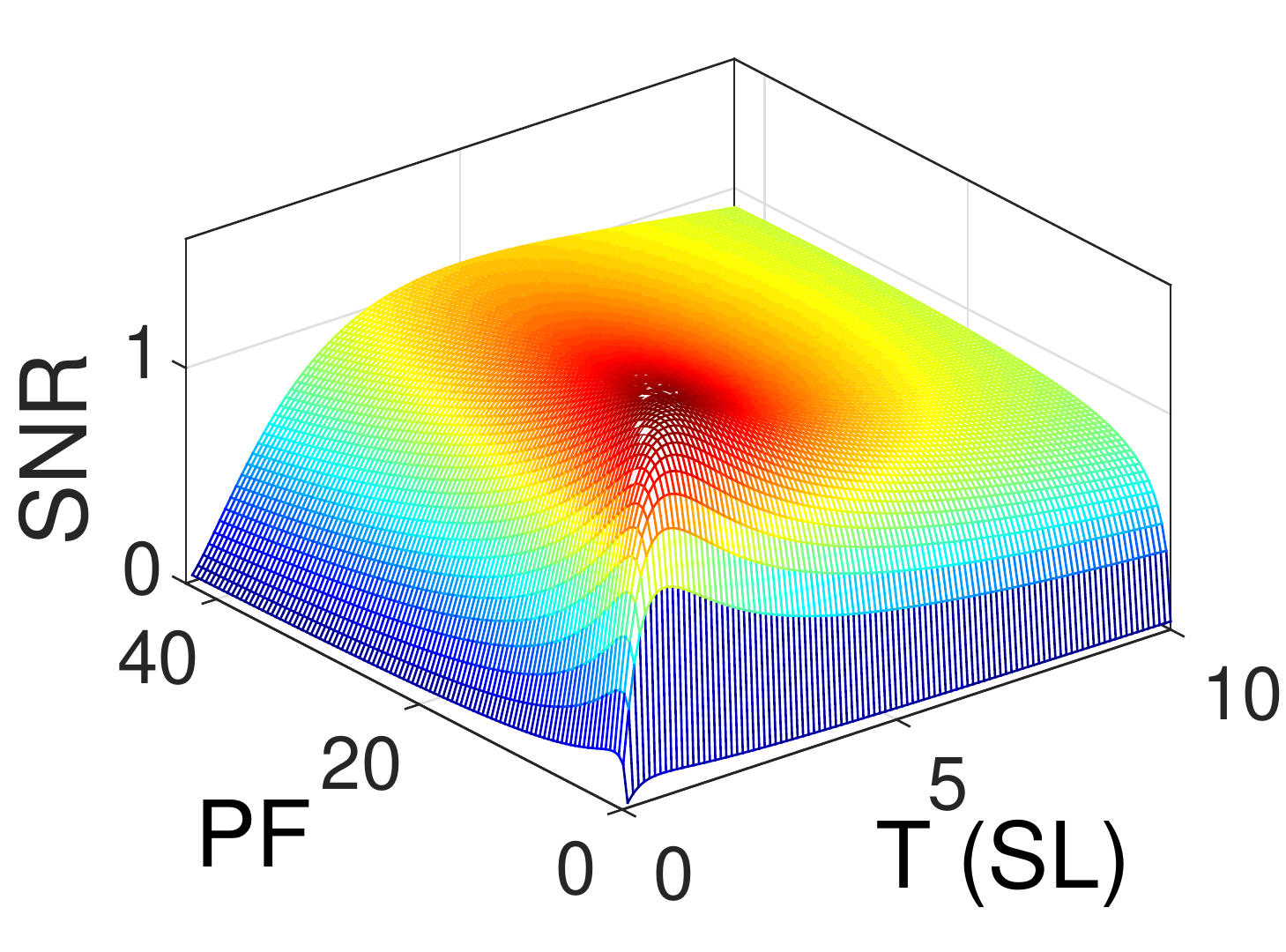}}
		\end{subfigure}
		\protect\caption{The effect of Purcell enhancement on the SNR. (a) SNR as a function of PF for typical rates and $T=1$. The SNR reaches a maximal value of $SNR=1.4867$ at the optimal Purcell factor $PF \approx 3$. The red dot shows the SNR for $PF=1$. (b) SNR as a function of PF and $T$ for typical rates. The SNR reaches a maximum value of $SNR=1.5951$ for optimal Purcell factor $PF=5.1903$ and pulse duration $T=1.703 [SL]$.} \label{fig:Kf_T}
	\end{center}
\end{figure}

When adding a resonant optical structure to the system, the radiative decay rate, $K_{f}$, can also be controlled. The Purcell factor can be defined by the ratio of the modified decay rate, $K_{f}$, to the original rate, $K_{f_0}$: $PF = K_{f}/K_{f_0}$. Increasing the radiative decay rate increases both $N_{0}$ and $N_{1}$, which have opposing effects on the SNR (since the contrast decreases). Figure \ref{fig:Kf_T}(a) depicts the SNR as a function of the Purcell factor, PF, with typical NV rates and $T=1 [SL]$. The plot clearly demonstrates that these conflicting effects give rise to a peak in the SNR that is achieved at an optimal PF. Specifically, the maximal $SNR=1.4867$ is reached at the optimal Purcell factor $PF \approx 3$. Figure \ref{fig:Kf_T}(b) depicts the SNR as a function of both $T$ and PF for typical NV rates, and demonstrates that optimizing the SNR over both PF and $T$ results in a slightly higher SNR of 1.5951 [compared to maximum SNR of 1.4867 with $T=1 [SL]$ in Figure \ref{fig:Kf_T}(a)], which is achieved for $PF \approx 5.1903$ and $T=1.703 [SL]$. This is a $\sim 27 \%$ increase compare to the maximum $SNR=1.2516$ with $PF=1$ [Fig. \ref{fig:T_Ke}(a)]. 

\begin{figure}[tbh]
	\begin{center}
		\begin{subfigure}[ ]{
				\includegraphics[trim = 1mm 1mm 10mm 2mm, clip, width=0.47 \columnwidth]{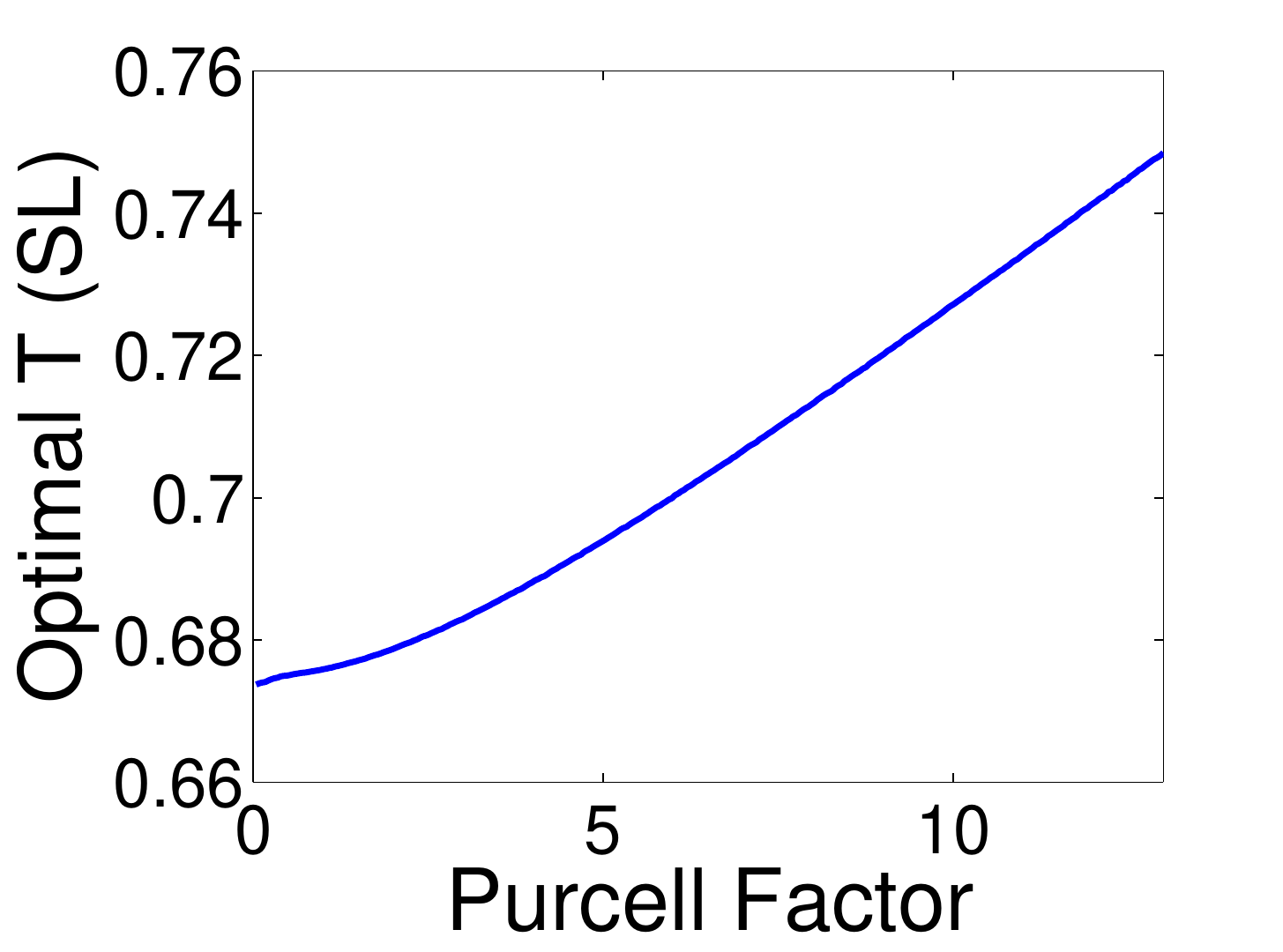}}
		\end{subfigure}
		\begin{subfigure}[ ]{
				\includegraphics[trim = 1mm 1mm 10mm 5mm, clip, width=0.47 \columnwidth]{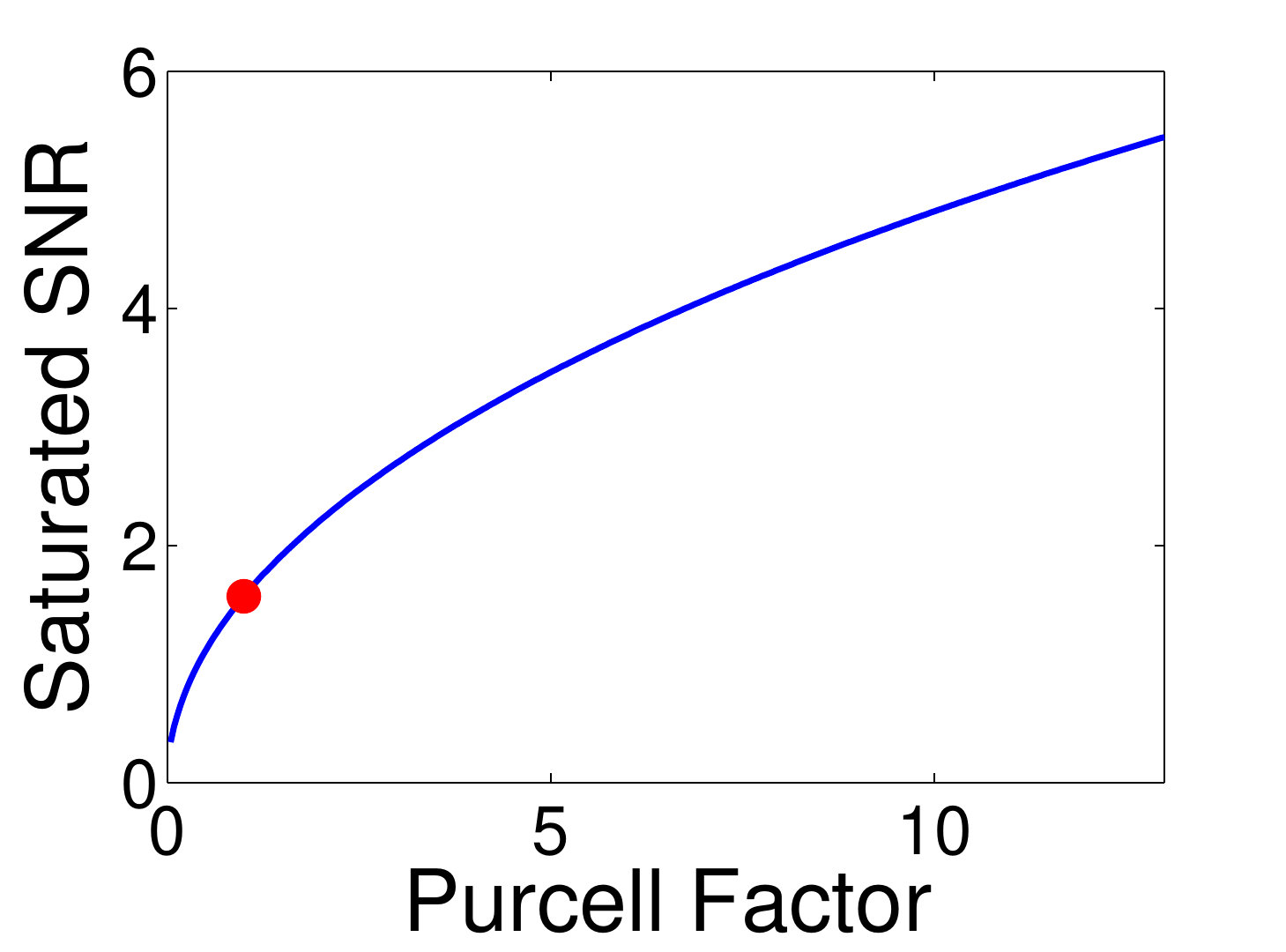}}
		\end{subfigure}
		\protect\caption{Saturated SNR for non-radiative spin-mixing transitions. (a) The optimal pulse duration, $T$, as a function of the PF for $K_e$ approaching infinity. (b) Saturated SNR as a function of the PF. The saturated value was	calculated by taking the optimal $T$ (for maximizing SNR) for each $K_e$  and calculating the maximal SNR when $K_e$ goes to infinity. The red dot shows the saturated SNR with $PF=1$. Increasing the PF and $K_e$ and adjusting $T$ to its optimal value can increase the SNR significantly.} \label{fig:PF_ExcitedMixing}
	\end{center}
\end{figure}

We observe that changing the excitation rate, $K_e$, or the Purcell factor alone does not increase the SNR considerably, as shown in figures \ref{fig:T_Ke} and \ref{fig:Kf_T}, respectively. However, one might expect that optimizing $T$, while taking $K_e$ to infinity (SNR saturation value) as well as increasing the PF, would take better advantage of the NV's saturation behavior and thus increase the SNR significantly. For a fixed Purcell factor, both the optimal $T$ (that maximizes the SNR) and the corresponding maximal SNR saturate when $K_e$ goes to infinity (see Fig. 1 in the Supplementary Material \cite{SuppMat}). This behavior allows us to investigate the effect of the Purcell factor on the asymptotic value of the maximal SNR (which will be referred to as the saturated SNR), and the corresponding asymptotic value of $T$ (which will be referred to as the optimal $T$). We note that choosing $K_e=2K_f=2PF*K_{f_0}$ approaches the saturation SNR values to within $15\%$. 

Fig. \ref{fig:PF_ExcitedMixing} depicts the saturated SNR as a function of PF for optimal $T$, assuming non-radiative spin-mixing rates that are not affected by the change in PF. The results clearly show a significant increase in the saturated SNR when increasing the PF, with no observable saturation. This demonstrates that under the assumptions that the spin mixing terms are of phononic origin (i.e. do not scale with the optical rates), the SNR can be improved significantly by increasing PF. This is the main result of this work. In particular, by reaching a moderate PF of only 4 the SNR can be doubled (compare to $PF=1$). Such a low PF should be easily achieved using broadband plasmonic nano-antennas for example, emphasizing the feasibility of such an SNR enhancement scheme. 

\begin{figure}[tbh]
\begin{center}
\begin{subfigure}[ ]{
\includegraphics[trim = 1mm 1mm 10mm 2mm, clip, width=0.47 \columnwidth]{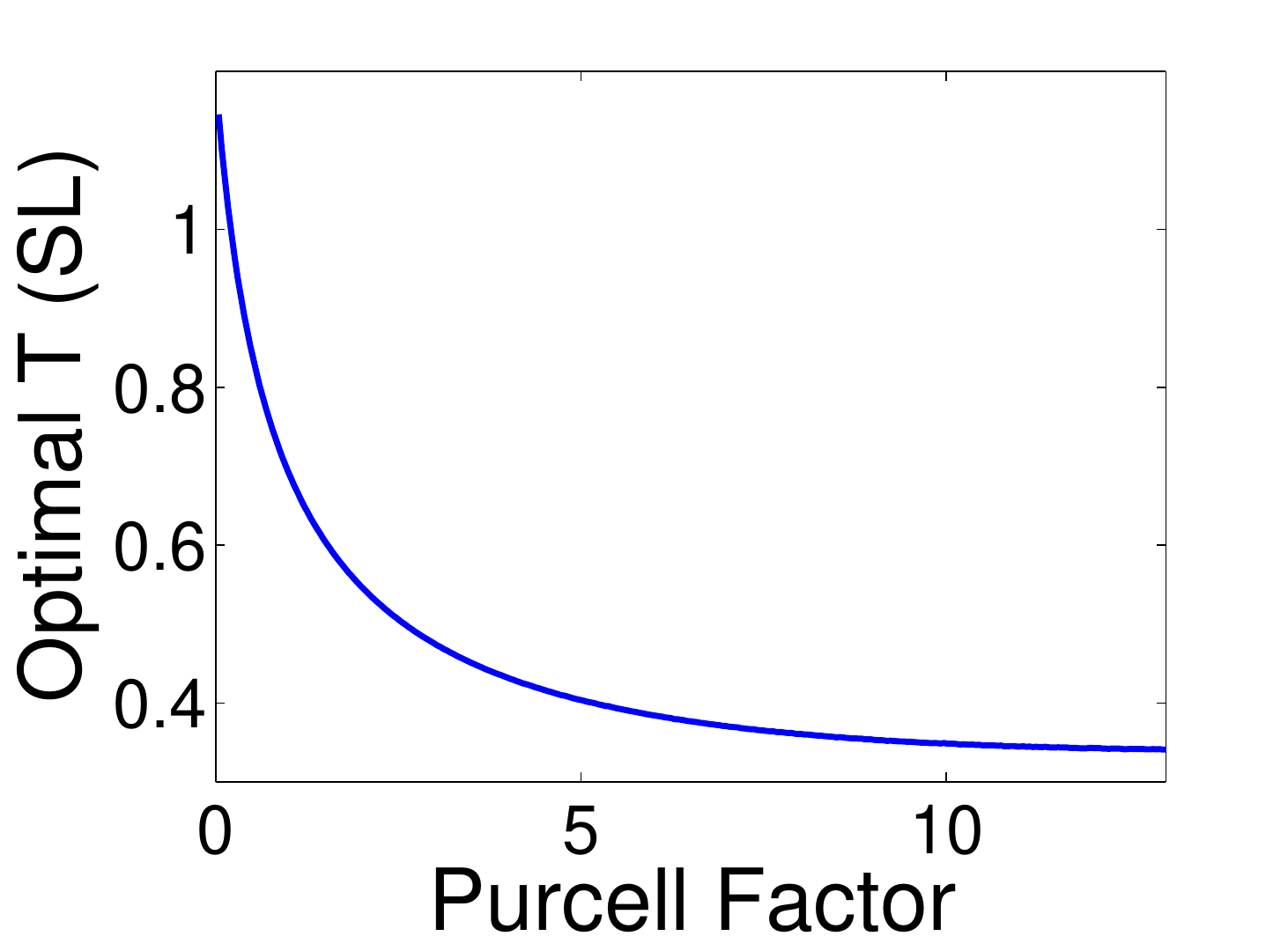}}
\end{subfigure}
\begin{subfigure}[ ]{
\includegraphics[trim = 1mm 1mm 10mm 6mm, clip, width=0.47 \columnwidth]{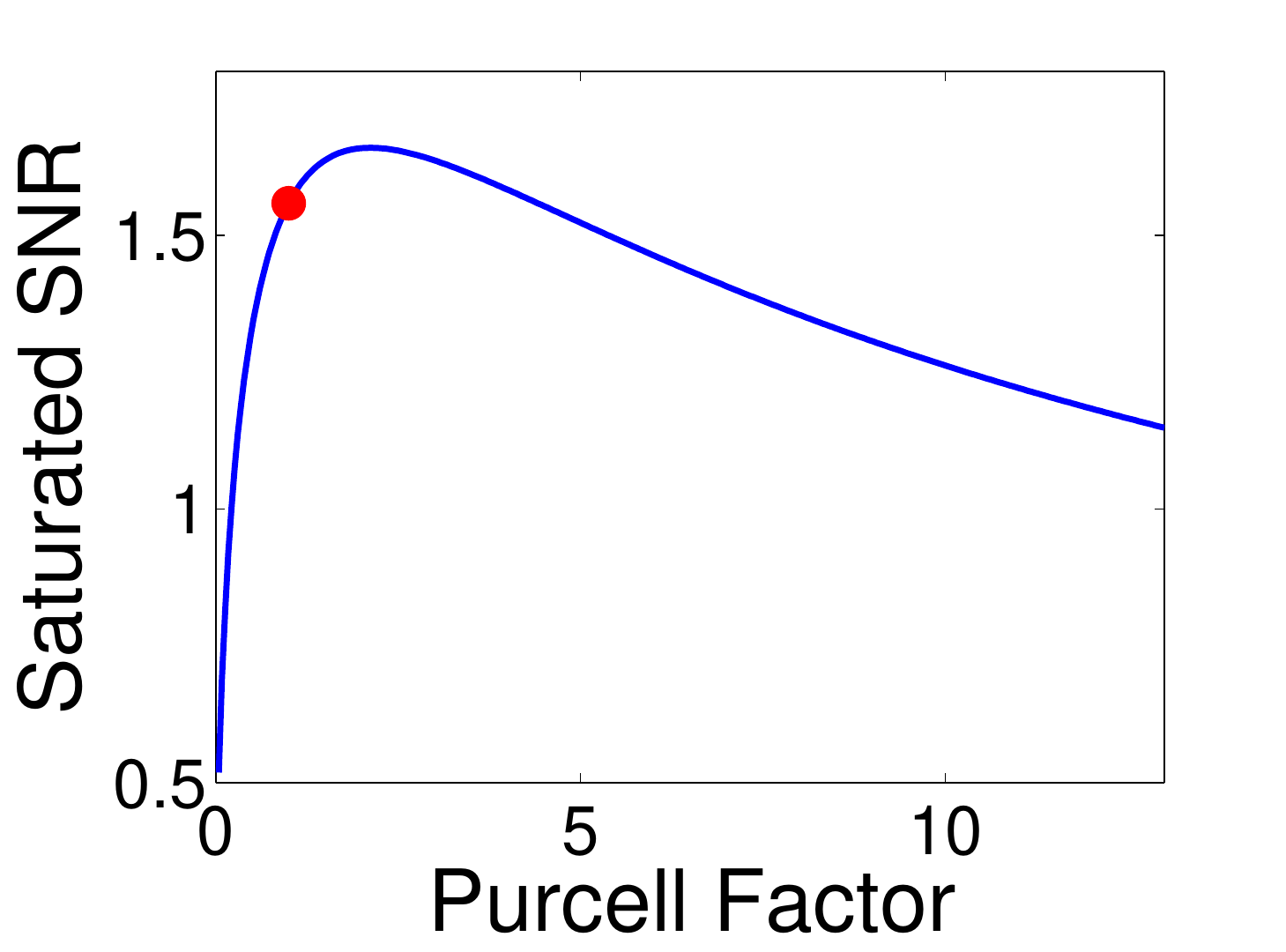}}
\end{subfigure}
		\protect\caption{Saturated SNR for radiative spin-mixing transitions. the mixing rates were set to be $K_{m_f}=0.02K_{f}\approx 0.4615PF$ and $K_{m_e}=0.02K_e$. (a) The optimal pulse duration, $T$, as a function of the PF for $K_e$ going to infinity. (b) Saturated SNR as a function of PF. The saturated value was calculated by maximizing the SNR over T as $K_e$ approaches infinity. The red dot marks the saturated SNR with $PF=1$. Optimizing $K_f$, $K_e$ and $T$ results in a maximal SNR value at the optimal $PF\approx 2$.} \label{fig:PF}
\end{center}
\end{figure}

We now analyze the opposite scenario, in which the spin-mixing terms are of radiative origin [Fig. \ref{fig:rates}(c)]. Figure \ref{fig:PF} shows that in this case the saturated SNR is bounded, and reaches its maximum at $PF\approx 2$. Optimizing the saturated SNR results only in a $\sim 6 \%$ improvement over the maximum SNR with the original decay rate ($PF=1$). The fact that the spin-mixing process is optically activated, and thus scales with $K_e$ and PF, strongly suppresses the SNR enhancement (since the spin contrast is limited).

We note that the measurements proposed in this work are highly sensitive to the relatively small spin-mixing terms. We find a significant (unsaturated) increase in the SNR assuming constant spin-mixing rates (Fig. \ref{fig:PF_ExcitedMixing}), while for radiative mixing rates (which scale with the Purcell factor) the results indicate only a moderate increase in the spin-readout SNR (Fig. \ref{fig:PF}). This strong dependence on the unknown origin of the spin mixing could provide a useful approach for studying the underlying physical processes related to this mixing at room temperature \cite{Doherty13}.

Changing the different rates of the system may also affect the optical initialization of the NV to $\ket{g,m_s=0}$. However, increasing the PF has only a minor effect on the ability to initialize the NV center. Increasing the PF to 10 causes a polarization decrease of only $\sim 1\%$ after initialization in the case of non-radiative spin-mixing, and a $\sim 6\%$ decrease for radiative spin-mixing rates.

In conclusion, we have analyzed the spin-readout SNR for NV centers in diamond, in the presence of optical couplers affecting their fluorescence rate through Purcell enhancement. We have constructed a relevant measure of the readout SNR, taking into account two competing processes - elevated signal yet reduced signal contrast - with increasing radiative decay rates. We find that the SNR can be improved by concurrently increasing both the decay and excitation rates, such that the NV saturation behavior is fully exploited. We find a significant and unsaturated improvement in the SNR as a function of the Purcell factor, assuming that spin-mixing rates are constant (unaffected by the Purcell enhancement). We note that this large improvement in readout SNR does not take into account potential improvements in collection efficiency that could result from the use of such optical couplers (e.g. \cite{haratz14,Choy13}), and could have a major impact on a variety of applications, potentially leading to single-shot spin readout. We also analyzed the system assuming that spin-mixing is radiative and scales with the Purcell factor. In this case the SNR is severely limited and saturates at $\sim 6 \%$ above its natural value (with $PF=1$). The strong dependence of our results on the origin of the spin-mixing terms suggests this approach as useful for studying the physics dominating such processes.

While preparing this manuscript we have become aware that related work is described in the thesis \cite{BabinecThesis}.

\begin{acknowledgments}
We thank Marko Loncar, Thomas Babinec, Phil Hemmer, Jorg Wrachtrup and Fedor Jelezko for fruitful discussions and useful insights. 
This work has been supported in part by the EU CIG, the Minerva ARCHES award, the Ministry of Science and Technology, Israel and the Israel Science Foundation (grant No. 750/14).
\end{acknowledgments}

\bibliography{NV}

\end{document}